\documentclass[runningheads]{llncs}

\usepackage[utf8]{inputenc}
\usepackage{algorithm}
\usepackage{algpseudocode}
\usepackage{amsmath}
\usepackage{hyperref}
\usepackage{booktabs}
\usepackage{graphicx}
\usepackage{amsfonts}
\usepackage{multirow}
\usepackage{subcaption}
\usepackage{diagbox}
\usepackage[misc,geometry]{ifsym}
\begin{document}
\title{E-SAGE: Explainability-based Defense Against Backdoor Attacks on Graph Neural Networks\thanks{This work was supported in part by the National Natural Science Foundation of China(NSFC) with Grant No. 62172383 and No. 62231015, Anhui Provincial Key R\&D Program with Grant No.S202103a05020098, Research Launch Project of University of Science and Technology of China(USTC) with Grant No.KY0110000049.}}
%

%
\author{Dingqiang Yuan\inst{1} \and
Xiaohua Xu\inst{1}\Letter \and
Lei Yu\inst{2}\and
Tongchang Han\inst{1} \and
Rongchang Li\inst{3} \and
Meng Han\inst{3} }
\authorrunning{D.Yuan et al.}
%
\institute{University of Science and Technology of China, Hefei 230000, China \email{dqyuan@mail.ustc.edu.cn, xiaohuaxu@ustc.edu.cn}\and
Rensselaer Polytechnic Institute,Troy NY 12180, USA 
\email{yul9@rpi.edu}\and
Zhejiang University, Hangzhou 310058, China
\email{mhan@zju.edu.cn}}

\maketitle              
\begin{abstract}
Graph Neural Networks (GNNs) have recently been widely adopted in multiple domains. Yet, they are notably vulnerable to adversarial and backdoor attacks. In particular, backdoor attacks based on subgraph insertion have been shown to be effective in graph classification tasks while being stealthy, successfully circumventing various existing defense methods. In this paper, we propose E-SAGE, a novel approach to defending GNN backdoor attacks based on explainability. We find that the malicious edges and benign edges have significant differences in the importance scores for explainability evaluation.
Accordingly, E-SAGE adaptively applies an iterative edge pruning process on the graph based on the edge scores. Through extensive experiments, we demonstrate the effectiveness of E-SAGE against state-of-the-art graph backdoor attacks in different attack settings. In addition, we investigate the effectiveness of E-SAGE against adversarial attacks.

\end{abstract}
\section{Introduction}
\vspace{-0.2cm}
Graph structures are widely used to model abstract relationships between entities in various domains \cite{wu2020comprehensive}\cite{zhang2022graph}. As a generalization of deep learning techniques for graph data, Graph Neural Networks (GNNs) have demonstrated powerful performance in areas such as recommendation systems \cite{fan2019metapath}, social networks \cite{guo2020deep}, and malware detection \cite{hei2021hawk}. However, GNNs are shown to be inherently vulnerable to malicious attacks \cite{zhou2021hierarchical}\cite{zhou2024reconstructed}. Adversarial attacks, due to their wide applicability and fewer restrictions, have garnered a lot of attention and led to the development of various attack and defense methods \cite{liu2018trojaning}\cite{Zou_2021}. Moreover, With the popularity of pre-trained models, public datasets, and federated learning \cite{tao2023byzantine,yu2024survey}, the possibility of embedding backdoors in pretrained models or sharing poisoned data has made more stringent backdoor attacks feasible \cite{yang2023percba}\cite{xu2023backdoor}\cite{yang2023persistent}. In backdoor attacks, attackers first poison the model during the training phase, and then, during the prediction phase, insert triggers into nodes or graphs that need to change the predictions, causing the target to be misclassified into the attacker’s desired category \cite{xu2021explainability}. Considering the feasibility, subgraph injection is currently the primary method for backdoor attacks. For example, in social networks, it is much easier to create several fake accounts and construct an adversarial subgraph than to log into the target account.

The explainability techniques of neural network models aim to determine the parts of the input features which play a key role in the prediction results. 
Integrated gradient \cite{sundararajan2016gradients,sundararajan2017axiomatic} is a classic method for quantifying the impact of input features. Recently Jiang et al. \cite{jiang2022defending} proposed a defense that leverages the explainability of GNNs to defend against backdoor attacks. However, this method has limitations, including the need for clean validation datasets and the potential for significant computational costs due to the calculation of explanatory metrics. 

To address these limitations, we propose a new explainability-based defense against backdoor in the node classification task. Our contributions are as follows: (i) We propose an adaptive edge pruning algorithm to prune adversarial subgraphs in the prediction stage. (ii) We extend the defense to address adversarial attacks and backdoor attacks that involve multiple subgraph insertions. To improve computation efficiency, we incorporate a neighbor sampling strategy akin to that used in GraphSAGE, leading us to name our approach E-SAGE\footnote{https://github.com/vanadisArya/E-SAGE}. We conducted extensive experiments on multiple datasets under various attack settings. Our experiments demonstrate the effectiveness of E-SAGE. 

The structure of the paper is organized as follows: we introduce the related work in Section 2. Section 3 elaborates on our proposed defense method, E-SAGE, including its design philosophy and implementation details. In Section 4, we present extensive experimental results to demonstrate the effectiveness and practicality of E-SAGE. Finally, Section 5 concludes the main contributions of this paper and discusses the potential directions of our future work.
\vspace{-0.2cm}
\section{Related Works}
\vspace{-0.2cm}
\subsection{Graph Neural Networks}
\vspace{-0.1cm}
Graph Neural Networks are pivotal for a variety of tasks such as node classification, graph classification, link prediction, and graph generation. This paper specifically delves into node classification, aiming to categorize target nodes by leveraging their attributes, the features of their neighbors, and the nodes' interconnections. Prominent models for this task include the Graph Convolutional Network (GCN) \cite{kipf2016semi}, GraphSAGE \cite{hamilton2017inductive}, and Graph Attention Network (GAT) \cite{velivckovic2017graph}. Among them, GraphSAGE improves efficiency by aggregating information from a limited number of neighbors.Conversely, GAT incorporates an attention mechanism to weigh the significance of each neighbor, enhancing model accuracy at the expense of increased computational demand and model complexity.
\vspace{-0.2cm}
\subsection{Graph Explainability and Integrated Gradients}
\vspace{-0.1cm}
Currently, two popular models for graph Explainability are GraphLIME \cite{huang2022graphlime} and GNNsExplainer \cite{ying2019gnnexplainer}. GraphLIME is a proxy-based Explainability method that extends the LIME \cite{ribeiro2016should} algorithm to GNNs and investigates the importance of different node features for node classification tasks. GNNsExplainer is a perturbation-based Explainability method that explores each node in the graph to identify the nodes and features that significantly impact the model's predictions. 
Integral Gradient \cite{sundararajan2016gradients}\cite{sundararajan2017axiomatic} is a method used to explain the predictions of deep neural network models. This method has been successfully used to explain the importance of certain pixels in convolutional neural networks(CNNs). Compared with other methods, the integral gradient method can more accurately determine which part of input are important for the output of the model.

\subsection{Subgraph-based attack and defense methods}
\vspace{-0.2cm}
\begin{figure}
\includegraphics[width=\textwidth]{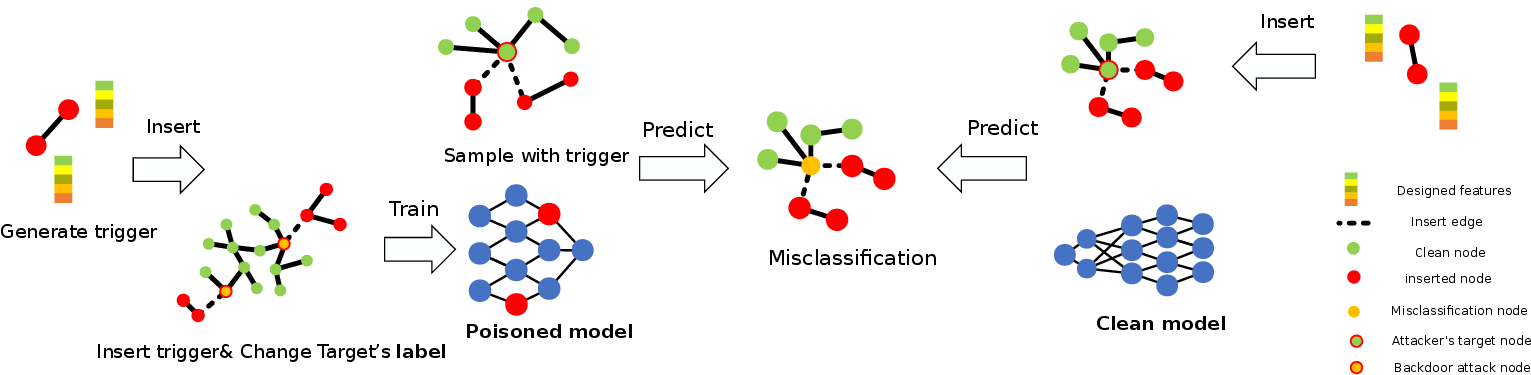}
\caption{Backdoor attack(left) and adversarial attack(right) model} \label{fig1}
\end{figure}
\vspace{-0.6cm}

\subsubsection{Subgraph-based attack}
According to the stage at which attacks occur, attacks on GNNs mainly focus on adversarial attacks and backdoor attacks. Fig.~\ref{fig1} gives an illustration of adversarial and backdoor attacks. Backdoor attacks typically involve attaching a backdoor trigger to the training data and assigning a target label to the samples with the trigger. If the backdoor is activated on a test sample that contains the embedded trigger, the model trained on the poisoned data will be misled. Zhang et al .\cite{zhang2021backdoor} proposed subgraph-based backdoor attack SBA on GNNs by injecting randomly generated universal triggers to some training samples. Xi et al .\cite{xi2021graph} adaptive triggers are generated for different graphs by calculating the similarity scores between nodes. Dai et al .\cite{dai2023unnoticeable} proposed UGBA which adaptive triggers are generated by utilizing the features of target nodes. Simultaneously, a hidden loss function is designed using cosine similarity between connected nodes to continuously optimize the trigger and resist defense methods. In addition, the most representative of graph adversarial attacks is Graph Injection Attack (GIA). GIA ensures that the node attributes and edges in the original graph $G$ are not modified while injecting new nodes or subgraphs into graph G to force the model to make incorrect predictions. Zou et al .\cite{Zou_2021} proposed TDGIA which according to the topological properties of the original graph $G$, important nodes are determined and new nodes are sequentially injected around them, and injects multiple subgraphs to attack.
\vspace{-0.4cm}
\subsubsection{Subgraph-based defense}
Prune is a simple yet practical defense mechanism. In this approach, if the cosine similarity between two nodes is low, the edge between them will be pruned. Since the edges created by backdoor attackers may link different nodes, the trigger structures and attached edges may be disrupted. Prune has been effective against existing backdoor attacks. However, the UGBA method designed by Dai et al . is optimized specifically against Prune, rendering Prune ineffective against UGBA. The defense method designed by Jiang et al .\cite{jiang2022defending} cannot be applied to situations where there is no original data and multiple subgraph insertions. Therefore, there is an urgent need for a defense method to resist the attacks mentioned above.
\vspace{-0.2cm}

\subsection{Threat Models}
\vspace{-0.1cm}
The backdoor attack based on subgraph insertion can be generally divided into two stages. Firstly, inserting subgraph into selected nodes and training a poisoned model with the poisoned data. During the prediction phase, a trigger is inserted to classify the target node into the category specified by the attacker. Therefore, the objective function of GNNs backdoor attacks can be formulated as follows:
\begin{equation}
    \left\{\begin{array}{l}h \circ f_\theta\left(m\left(G ; g_t\right)\right)=y_t \\ h \circ f_\theta(G)=h \circ f_{\theta_0}(G)\end{array}\right.
\end{equation}
Note: $h$ is the downstream classifier after fine-tuning. $G$ represents the given graph data, $g_t$ represents the trigger, $m\left(G ; g_t\right)$ represents the graph data fused with the trigger, $f_\theta(G)$ represents the backdoored GNNs model, and $f_{\theta_0}(G)$ represents the GNNs model without a backdoor. The first equation indicates that the graph data with the trigger will produce the desired results as intended by the attacker. The second equation states that for data without the trigger, the output will be the same as that of the GNNs model without a backdoor. 

\section{E-SAGE Defense}
In this chapter, we first introduce the capabilities and objectives of the defender, followed by the design rationale of our defense method, and finally, we present the corresponding algorithm.
\vspace{-0.2cm}
\subsection{Defense overview}
\vspace{-0.1cm}
\subsubsection{Attacker's goal and capabilities}
For backdoor attacks, the attacker can train the model using poisoned data and inject adversarial subgraphs into the target nodes in the prediction phase. The attacker needs to maintain model accuracy when samples do not contain triggers. When the input sample contains a trigger, the model misclassifies it into the classification specified by the attacker.
\vspace{-0.5cm}
\subsubsection{Defender's goal and capabilities}
The defender's goal is to ensure the accuracy (ACC) of the model and reduce the attacker's attack success rate (ASR).The defender usually does not know whether the model is poisoned, so adversarial edges between triggers and target nodes need to be pruned during the prediction phase.
\vspace{-0.2cm}
\subsection{Defense design}
\vspace{-0.1cm}
Our defense method is founded on three key pillars, which encompass its design principles and underlying rationale:
\vspace{-0.4cm}
\subsubsection{Integral gradient based explainer}
Subgraph-based attacks aim to alter the target node's prediction with a minimal budget, significantly impacting the model's classification results, whether through adversarial or backdoor tactics. This results in a large change in the predicted values before and after adversarial subgraph insertion for the model. After quantifying this change through the integral gradient in Equation~\ref{eq2}, we can obtain the importance score of each edge for prediction.
\begin{equation}\label{eq2}
\operatorname{IntegratedGrads}_i(x)=\left(x_i-x_i^{\prime}\right) \times \int_{\alpha=0}^1 \frac{\partial F\left(x^{\prime}+\alpha \times\left(x-x^{\prime}\right)\right)}{\partial x_i} d \alpha
\end{equation}
Here, $x$ represents the input sample, $x^{\prime}$ is a baseline input (typically a zero input or an average input), $F$ is the graph neural network model. In GNNs, we input both the information of edges and node features into the explainer to obtain an analysis of the importance of features and edges.Specifically, we convert all edges into edge mask tensors with a value of 1, thus becoming a continuously transformable input. In this way, we can obtain the importance of edge $x_i$ for prediction.

\begin{figure}[htp]
\centering
\begin{subfigure}[b]{0.47\textwidth}
  \centering
  \includegraphics[width=\linewidth]{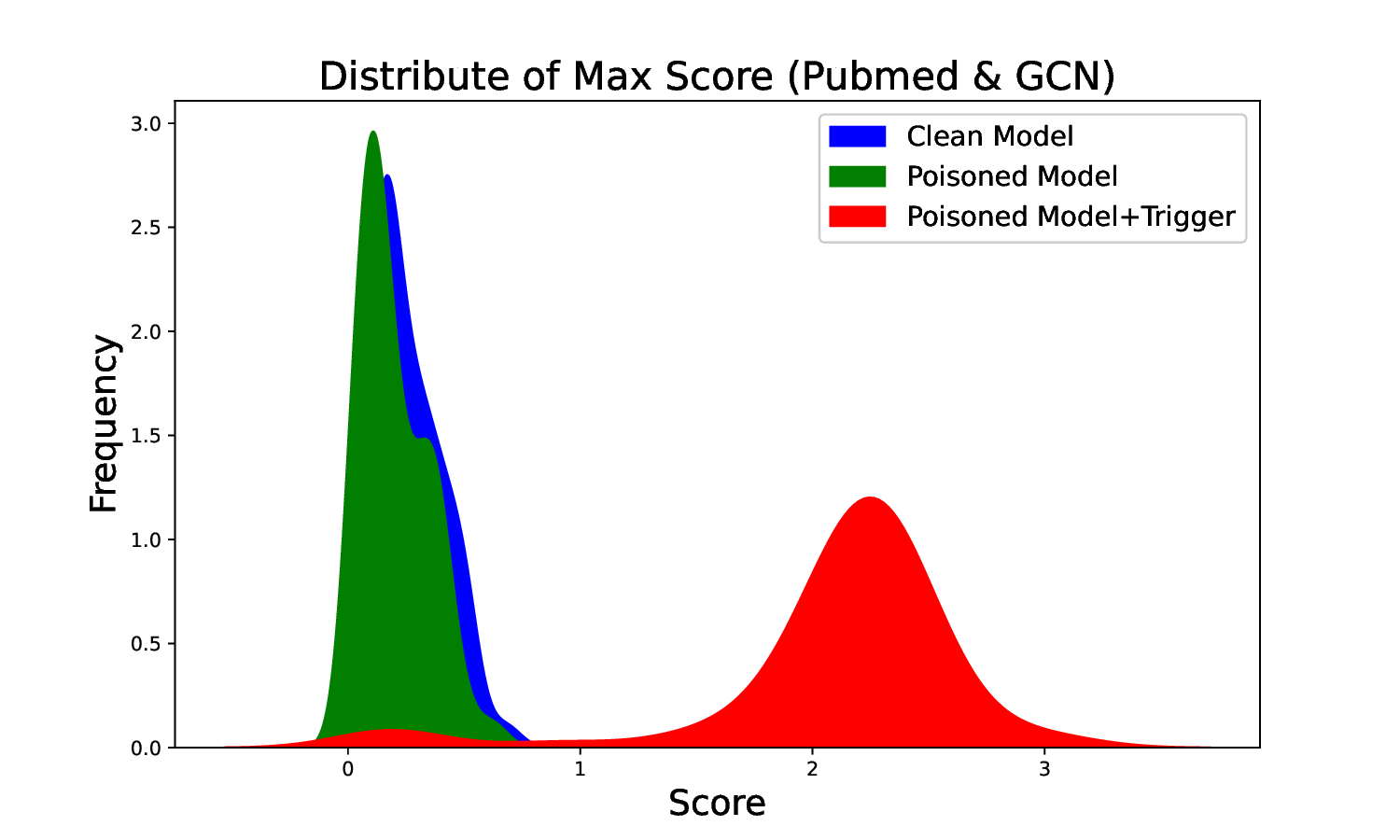}
\end{subfigure}
\hfill
\begin{subfigure}[b]{0.47\textwidth}
  \centering
  \includegraphics[width=\linewidth]{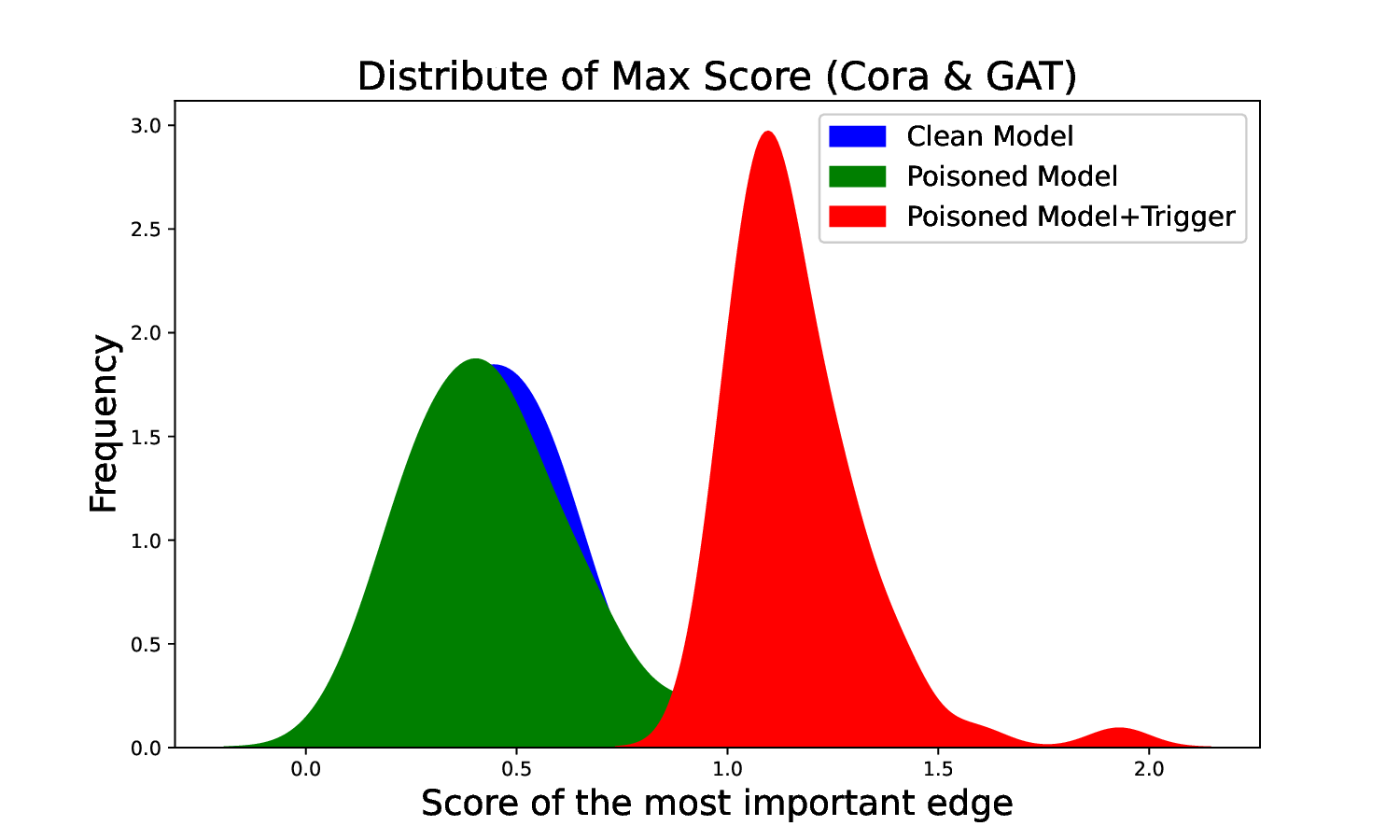}
\end{subfigure}
\hfill
\begin{subfigure}[b]{0.47\textwidth}
  \centering
  \includegraphics[width=\linewidth]{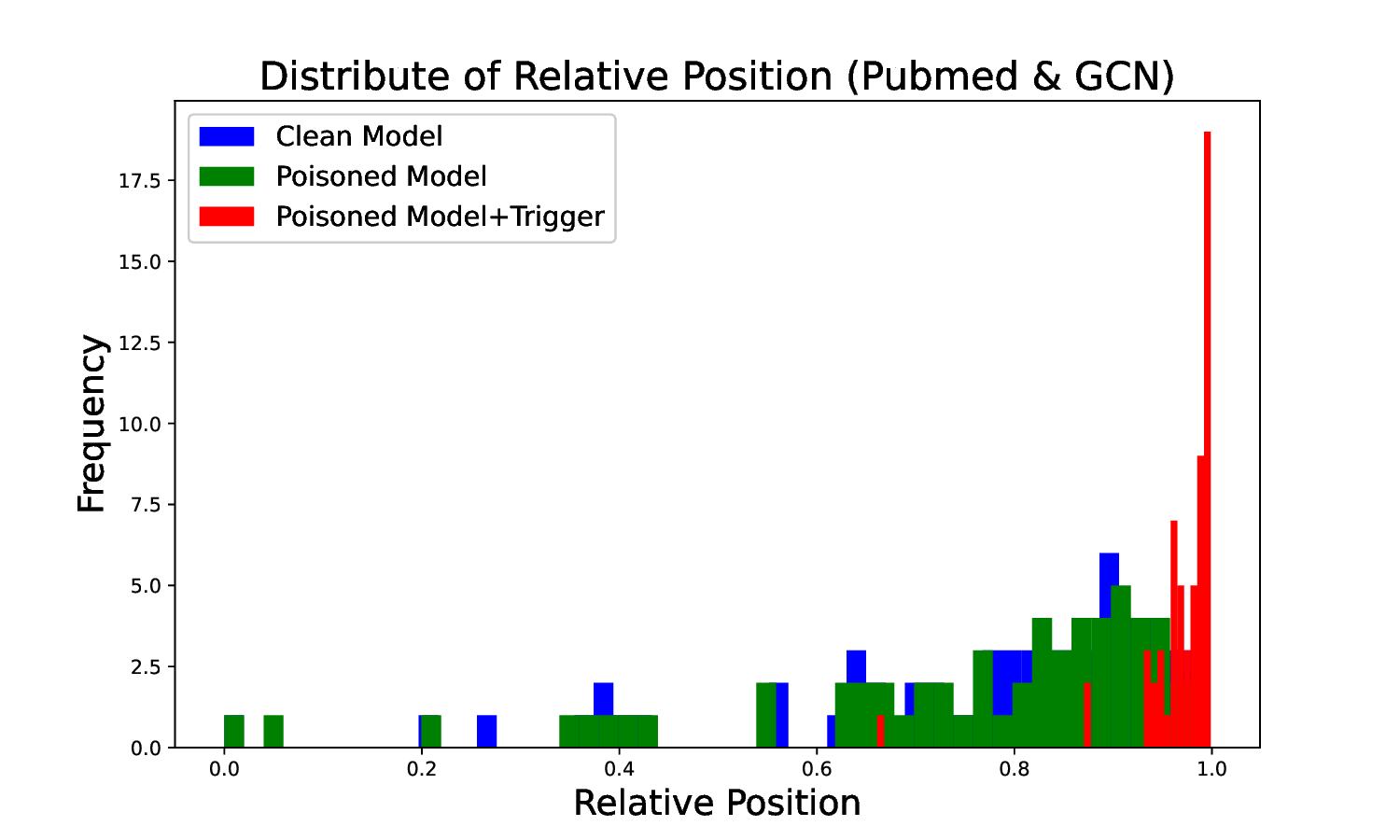}
\end{subfigure}
\hfill
\begin{subfigure}[b]{0.47\textwidth}
  \centering
  \includegraphics[width=\linewidth]{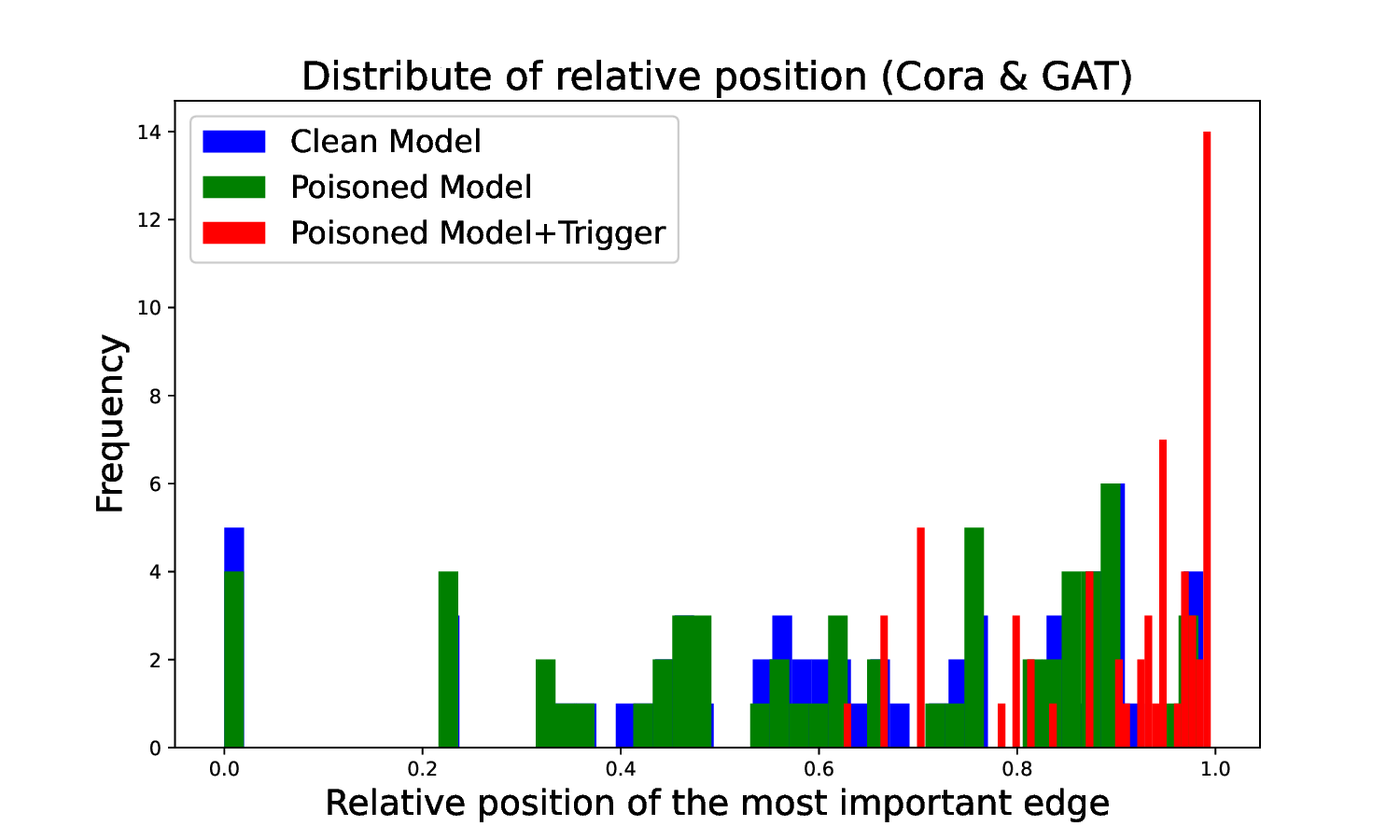}
\end{subfigure}
\hfill
\caption{Score and Position Distribution of the Most Important Edges}
\label{fig2}
\end{figure}

Next, we show the results given by the explainer in different cases. The results shown in Fig.~\ref{fig2} indicate that when no trigger is inserted into the test subgraph, the poisoned model exhibits similar performance to the normal model. However, once a trigger is inserted, the scores of the most important edges in the sample rise sharply and their relative positions are concentrated towards the end of the edge set. This proves that the most important edge is the adversarial edge we inserted at the end of the set.

Now, we need to determine a threshold \(\beta\) and prune edges that exceed this value. The distribution in Fig.~\ref{fig2} indicates that there is a broader range of choices for 
\(\beta\) in larger datasets. Yet, in smaller datasets, there exists a trade-off between accuracy and defensive performance. Specifically, if \(\beta\) is too small, it may lead to the removal of clean edges, thereby reducing accuracy, which is particularly evident in nodes with lower degrees. Conversely, if \(\beta\) is too large, it may fail to prune adversarial subgraphs. Therefore, we use the sigmoid function to control \(\beta\) within the range of 0.45 to 0.8, decreasing it with an increase in degree, ensuring a more lenient boundary when the degree is low. 
\vspace{-0.5cm}
\subsubsection{Iteration}
\begin{figure}[htp]
\centering
\begin{subfigure}[b]{0.45\textwidth}
  \centering
  \includegraphics[width=\textwidth]{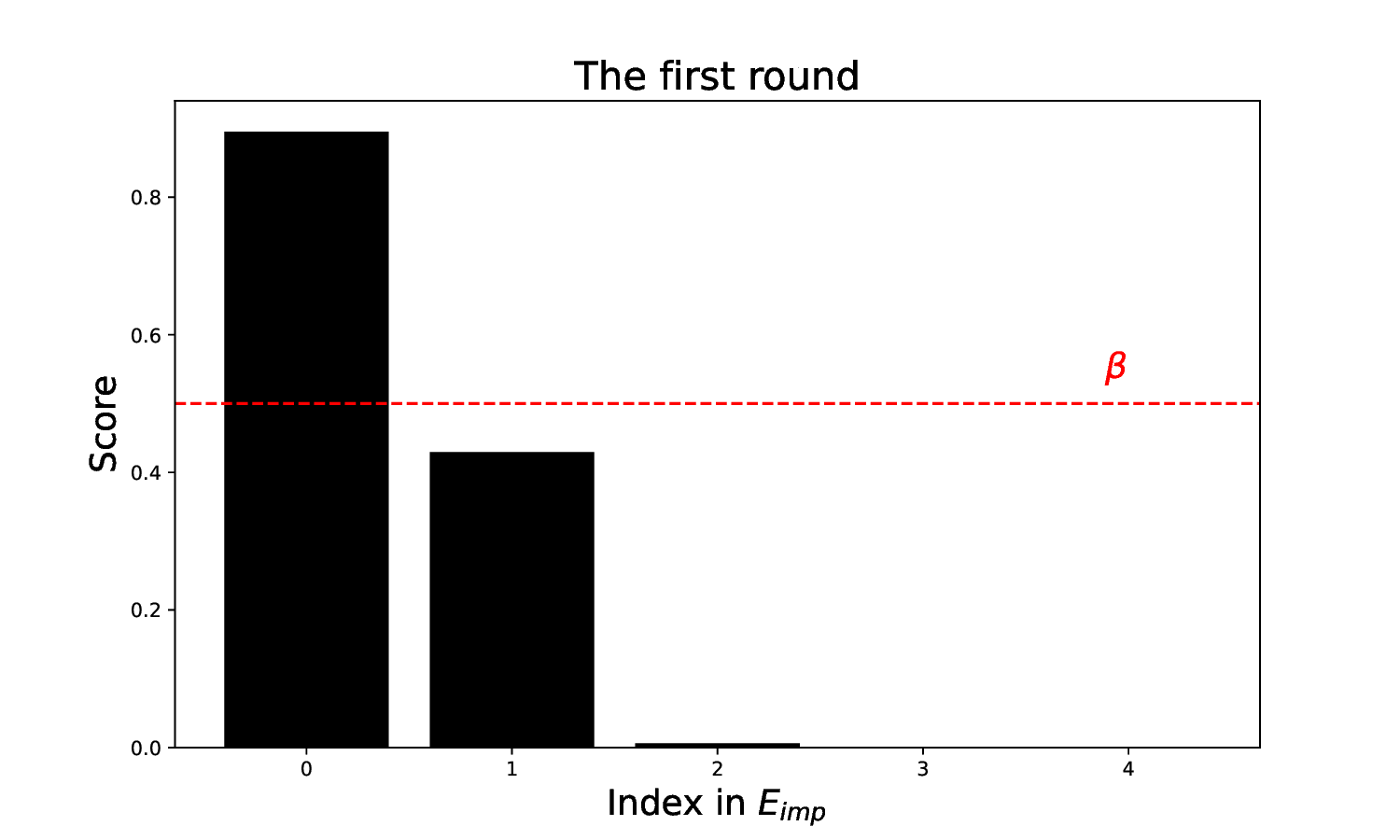}
\end{subfigure}
\hfill
\begin{subfigure}[b]{0.45\textwidth}
  \centering
  \includegraphics[width=\textwidth]{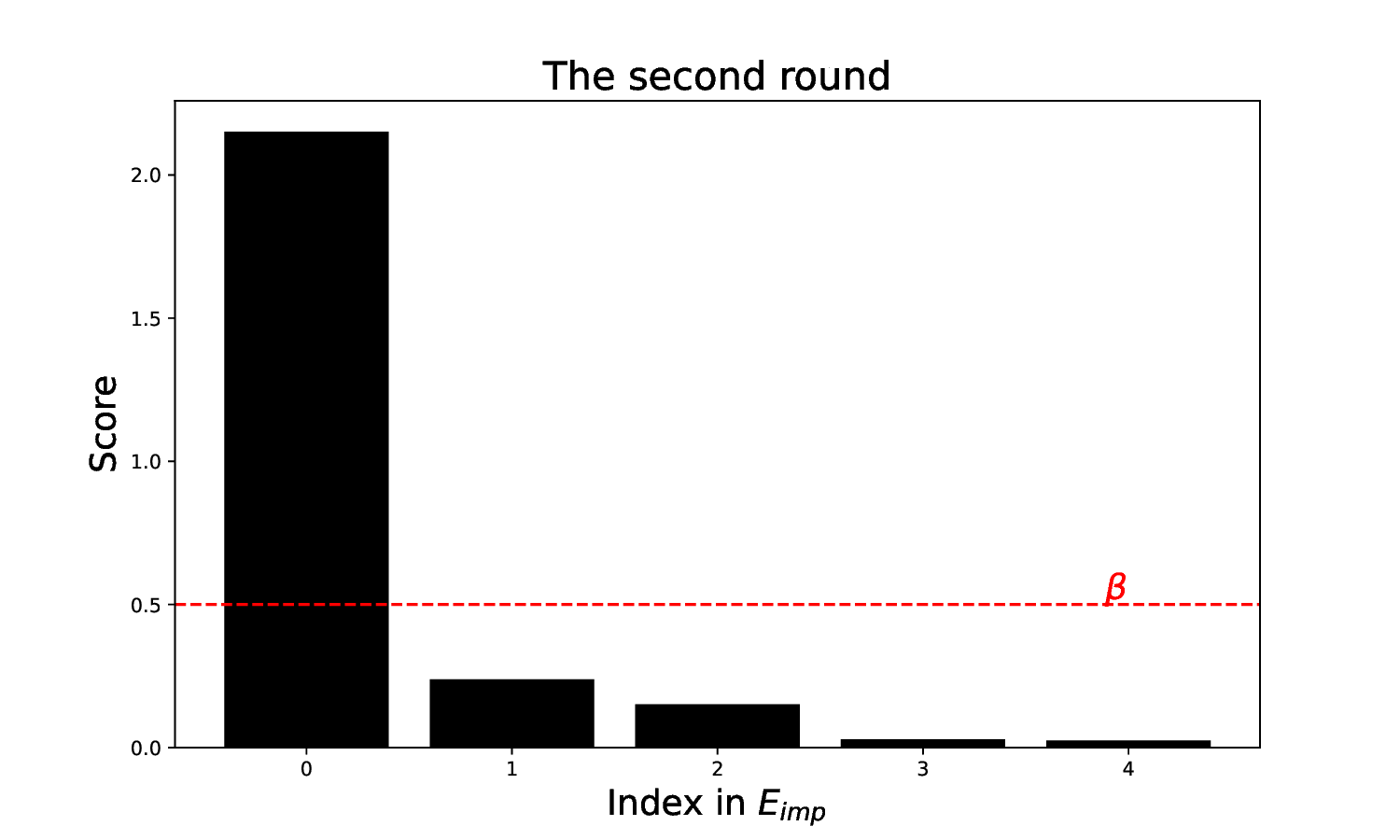}
\end{subfigure}
\hfill

\begin{subfigure}[b]{0.45\textwidth}

  \centering
  \includegraphics[width=\textwidth]{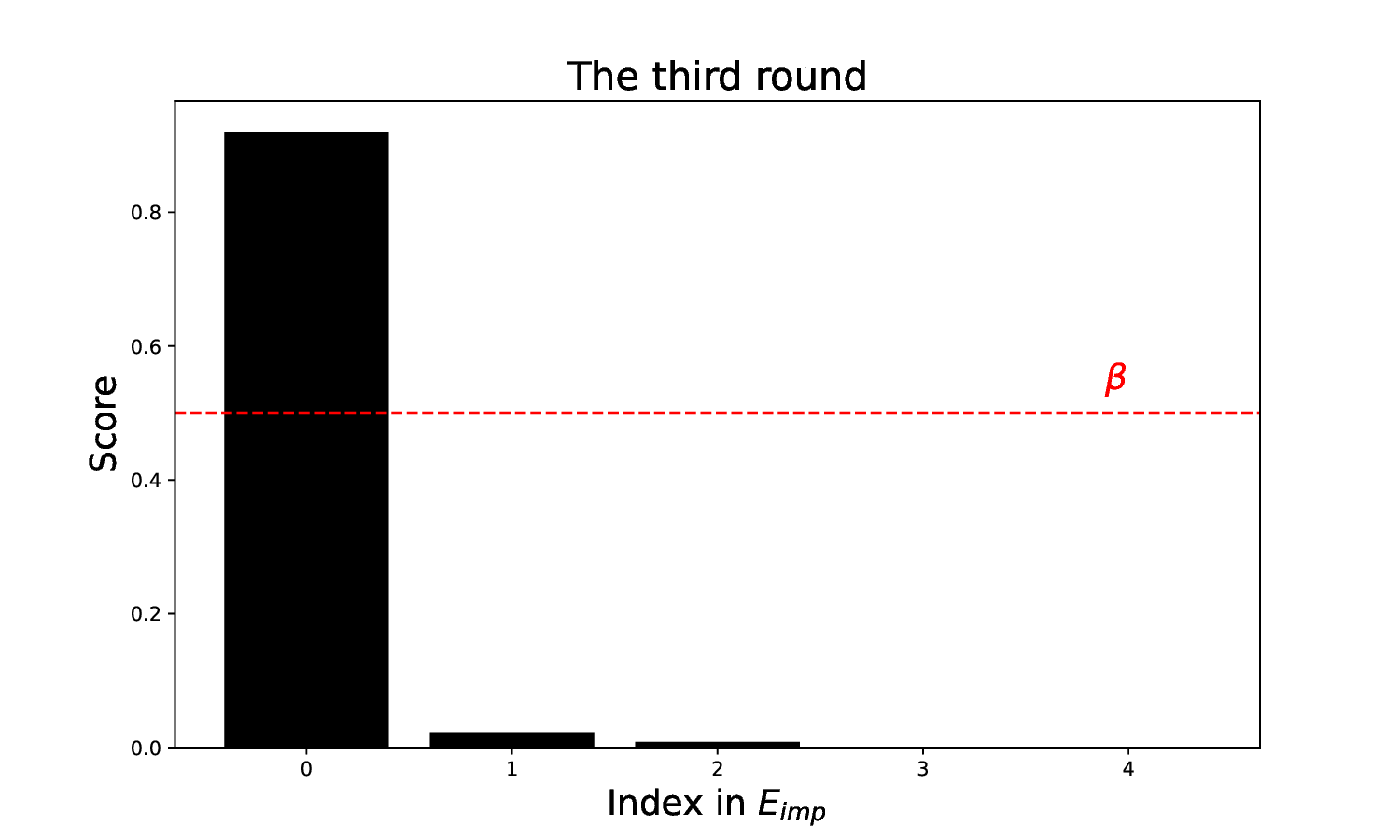}
\end{subfigure}
\hfill
\begin{subfigure}[b]{0.45\textwidth}
  \centering
  \includegraphics[width=\textwidth]{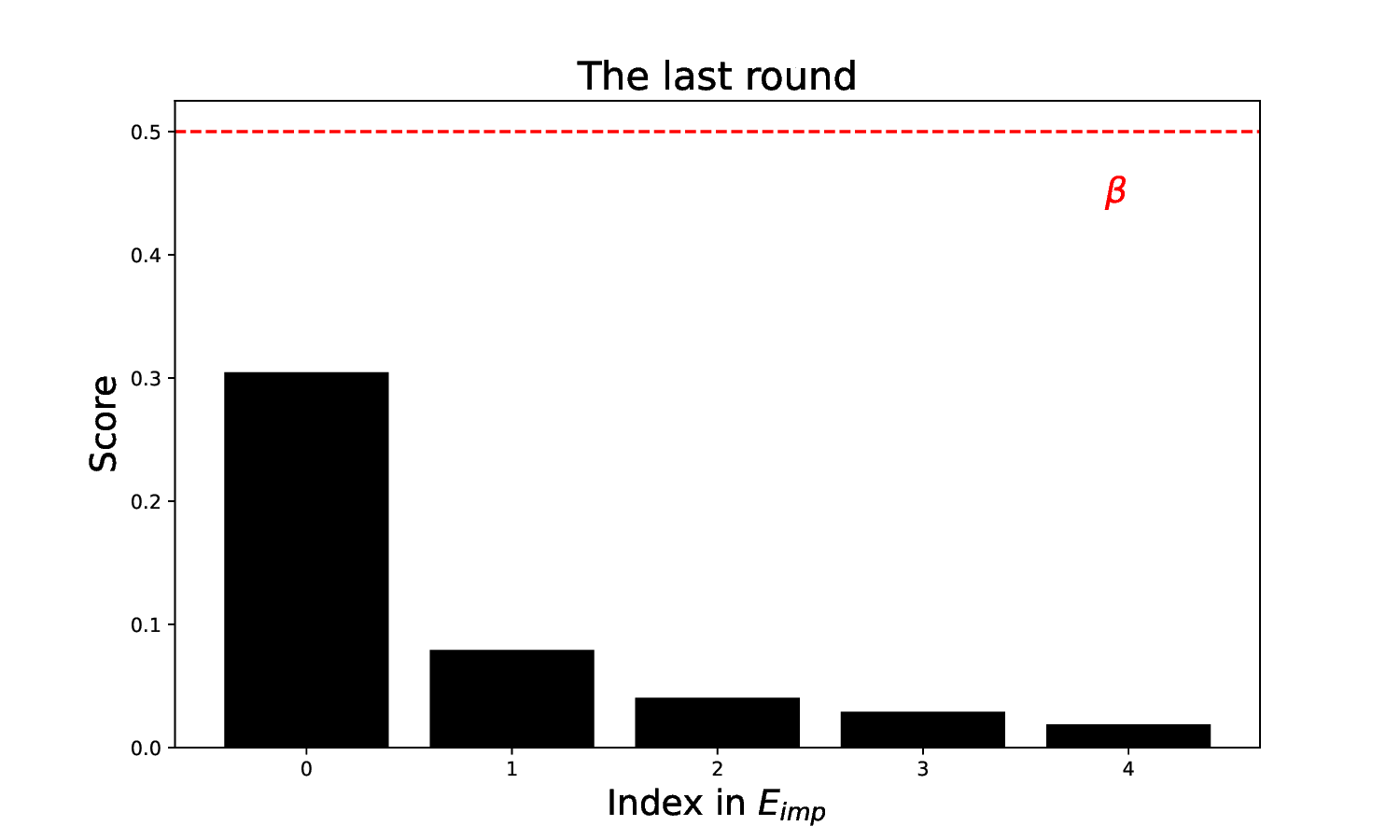}
\end{subfigure}
\hfill

\caption{Iteration of inserting 3 subgraphs in TDGIA attack}
\label{fig3}
\end{figure}

Adversarial attacks based on subgraph insertion are similar in behavior and objectives to backdoor attacks. In principle, our method should also be applicable to adversarial attacks. TDGIA can insert any number of adversarial subgraphs into the target. In testing the defense against TDGIA, we found that a single round of pruning could not defend against attacks involving the insertion of multiple subgraphs. The reason is that some critical adversarial edges make others seem less important. This suggests that if attackers employ such a method in backdoor attacks, our defense method would be obscured. Therefore, we prune the samples in an iterative manner. Fig.~\ref{fig3} shows the iterative pruning process for samples with three inserted adversarial edges and how some adversarial edges can obscure others.
\vspace{-0.5cm}
\subsubsection{Simplifying Input Samples}
We are well aware that in node prediction tasks, the average degree of nodes is usually high. As the radius of the subgraph increases, the number of nodes and edges will increase exponentially. Therefore, we have opted for an approach similar to GraphSAGE, which involves discarding some nodes with lower relevance in exchange for improved efficiency. For each neighbor \(N_i\) of the target node \(N_t\), we only randomly retain its \(\theta_d\) neighbors.
\vspace{-0.2cm}
\subsection{Defense method}
\vspace{-0.1cm}
Algorithm~\ref{Alg1} demonstrates the operational logic of our defense method. Firstly, we input the graph neural network \( f \) required for prediction,  the graph \( G_p(V,E,X) \), the target node \( N_t \in V\), and the threshold parameter \( \theta_e \) for subgraph edges, as well as the threshold parameter \( \theta_n \) for retaining neighbors when simplifying the graph. This process is also called neighbor sampling. Lines 4 to 12 of the algorithm illustrate our simplification logic, which is essentially to simplify the input graph when there are too many edges, retaining only a limited number of nodes \( V' \) and edges \( E' \). The algorithm demonstrates the procedure when the graph radius is 2, where \( \mathcal{N}(N_i) \) represents the set of neighboring nodes of node \( N_i \). From lines 13 to 20, we obtain ordered arrays of scores\( S_{\text{imp}} \) and edge indices\( E_{\text{imp}} \) through the explainer \( Expl \). If the score of the most important edge exceeds the threshold \( \beta \), then we prune it. The function that determines \(\beta\) is a variant of the sigmoid function:\[\text{sigmoid}^*(\text{deg}) = 0.45 + 0.35 \times \left(1 + e^{1.2(\text{deg} - 3.5)}\right)^{-1}\] We then iterate this process until reaching the pruning limit \( cut_\text{max} \)  or there are no suspicious edges left.
\begin{algorithm}
\caption{:E-SAGE Defense}\label{Alg1}
\begin{algorithmic}[1]
\State \textbf{Input:}  Model \( f \), \( G_p(V,E,X) \), Target node \( N_t \)
\State \textbf{Output:} Pruned Edge Set \( E' \)
\State \textbf{Initialize:} \( max\_cut = \max(deg(N_t)/4, 1) \), \( cut = 0 \), \( \beta = sigmoid^*(deg(N_t)) \)
\If {\( |E| > \theta_{\text{e}} \)} 
    \For{each \( N_i \in \mathcal{N}(N_t) \ \)}
    \State  \( \mathcal{N}(N_i) \gets \text{Randomly select } \theta_{\text{n}} \text{ neighbors from } \mathcal{N}(N_i)\setminus N_t \)
        \State Add \( N_t \) to \( \mathcal{N}(N_i) \)
    \EndFor
    \State Obtain new subgraph \( G'_p(V', E', X') \)
\Else
    \State \( G'_p = G_p \)
\EndIf

\While{\( cut < max\_cut \)}  
    \State \( (S_{\text{imp}}, E_{\text{imp}}) \gets \text{Expl}(f,G'_p, N_t) \) 
    \If{\( S_{\text{imp}}[0] < \beta \)}
        \State \textbf{break}
    \EndIf
    \State \( E' = E' \setminus \{E_{\text{imp}}[0]\} \)  
    \State \( cut += 1 \)
\EndWhile
\State \textbf{return} \( E' \)
\end{algorithmic}
\end{algorithm}
\vspace{-0.2cm}
\section{Experiment}
\vspace{-0.2cm}
In this chapter, we verify the effectiveness of our defense method by conducting experiments on various datasets and GNNs models. (i) We applied our method to defend against backdoor attacks from SBA, GTA, and UGBA. (ii) We have expanded the attack scope to include attacks based on the insertion of multiple subgraphs (e.g., TDGIA). (iii) We evaluated the time consumption of the algorithm.
\vspace{-0.2cm}
\subsection{Experimental settings and metrics}
\vspace{-0.1cm}
We conducted experiments on three mainstream models: GCN, GAT, and GraphSAGE. The experiments primarily utilized datasets of three different sizes, namely Cora, Pubmed, and Flickr, with their basic data presented in Table.~\ref{tab1}. Unless specifically noted, the results are derived from the mean of three models. The primary computational device used in our experiments is the NVIDIA GeForce RTX 4090. This work was supported by frameworks such as Captum, UGBA and pytorch.
\vspace{-0.2cm}
\begin{table}[h]
\centering
\setlength{\tabcolsep}{12pt}
\renewcommand{\arraystretch}{1.25}
\caption{Dataset Statistics}
\begin{tabular}{lccccc}
\toprule
Dataset & Nodes & Features & Classes & Edges & Average Degree \\
\midrule
Cora    & 2708            & 1433               & 7                 & 10556           & 3.9            \\
Pubmed  & 19717           & 500                & 3                 & 88648           & 4.5            \\
Flickr  & 89250           & 500                & 7                 & 899756          & 10.1           \\
\bottomrule
\end{tabular}
\label{tab1}
\end{table}

In the experiments comparing attack and defense methods, we uniformly set the number of adversarial subgraphs that can be inserted to be 0.5\% of the test set. The algorithm's parameters were uniformly set to \(\theta_{\text{node}}=5\), \(\theta_{\text{edge}}=300\), with the attack method's \(\text{trigger\_size}=3\), and the number of layers in the graph neural networks as 2. Our measurement parameters included the ASR and ACC. ACC represents the accuracy of the model. The ASR is given by the formula:
\[ \text{ASR} = \frac{|\{v \in V_{\text{test}} : f_p(v, G) = y'\}|}{|V_{\text{test}}|} \]
We use \(f_p\) represents the poisoned model, and \(y'\) is the attacker's target class.
\vspace{-0.2cm}
\subsection{Defense Result}
\vspace{-0.1cm}
We defend against existing backdoor attacks on multiple models and datasets. As shown in Table~\ref{tab2}, our method maintains high ACC while significantly reducing ASR. This shows that our defense method has achieved good results in facing existing attacks. Especially for UGBA that other defense methods cannot handle, our explainability-based method can accurately find out its inserted triggers.
\vspace{-0.2cm}
\begin{table*}
\setlength{\tabcolsep}{12pt}
\renewcommand{\arraystretch}{1.25}
\centering
\caption{Backdoor defense result(ASR($\%$)\(|\)ACC($\%$))}
\begin{tabular}{c  c   c   c   c}
\hline {Dataset} & {Method} &{None} & {Prune} & {Ours}  \\
\hline
 
\multirow{3}{*}{Cora} 

& {SBA-Gen} & {37.66}\(|\)57.21 & {8.21}\(|\)78.87 & \textbf{7.82}\(|\)78.22  \\
& {GTA} & {95.04}\(|\)4.44 & {6.74}\(|\)79.56 & \textbf{4.74}\(|\)79.56  \\
& {UGBA} & {80.13}\(|\)15.14 & {75.83}\(|\)21.50 & \textbf{6.06}\(|\)80.00 \\

\hline
 
\multirow{3}{*}{Pubmed} 

& {SBA-Gen} & {50.28}\(|\)46.98 & {8.80}\(|\)84.62 & \textbf{4.28}\(|\)87.93  \\
& {GTA} & {90.03}\(|\)9.10 & {4.32}\(|\)88.16 & \textbf{4.47}\(|\)87.07  \\
& {UGBA} & {93.20}\(|\)6.23 & {92.90}\(|\)6.43 & \textbf{6.39}\(|\)84.39 \\

\hline
 
\multirow{3}{*}{Flickr} 

& {SBA-Gen} & {0.00}\(|\)50.20 & {0.00}\(|\)51.23 & {0.00}\(|\)50.07  \\
& {GTA} & {98.61}\(|\)0.67 & {6.86}\(|\)45.10 & \textbf{0.51}\(|\)47.37  \\
& {UGBA} & {97.63}\(|\)2.03 & {97.73}\(|\)1.90 & \textbf{0.07}\(|\)49.50 \\

\hline
 
\hline \end{tabular}
\label{tab2}
\end{table*}

 We have deployed our defense method on the clean model and conducted experiments. Table~\ref{tab3} shows that, except for the small dataset Cora, our defense does not cause a loss of accuracy.
\begin{table*}
\setlength{\tabcolsep}{12pt}
\renewcommand{\arraystretch}{1.25}
\centering
\caption{Accuracy(ACC($\%$))  on Clean models}\label{tab3}

\begin{tabular}{c  c    c  c}
\hline {\diagbox{Dataset}{Defense}} & {none} &{Prune}&{Ours} \\
\hline

{Cora} & {82.88} & {80.25} & {77.65}  \\
\hline
{Pubmed} & {84.44} & {83.84}  & {84.53} \\
\hline
{Flickr} & {41.80} & {41.00} & {41.80} \\

\hline \end{tabular}
\label{tab3}
\end{table*}

\begin{figure}[htp]
\centering
\begin{subfigure}[b]{0.47\textwidth}
  \centering
  \includegraphics[width=\linewidth]{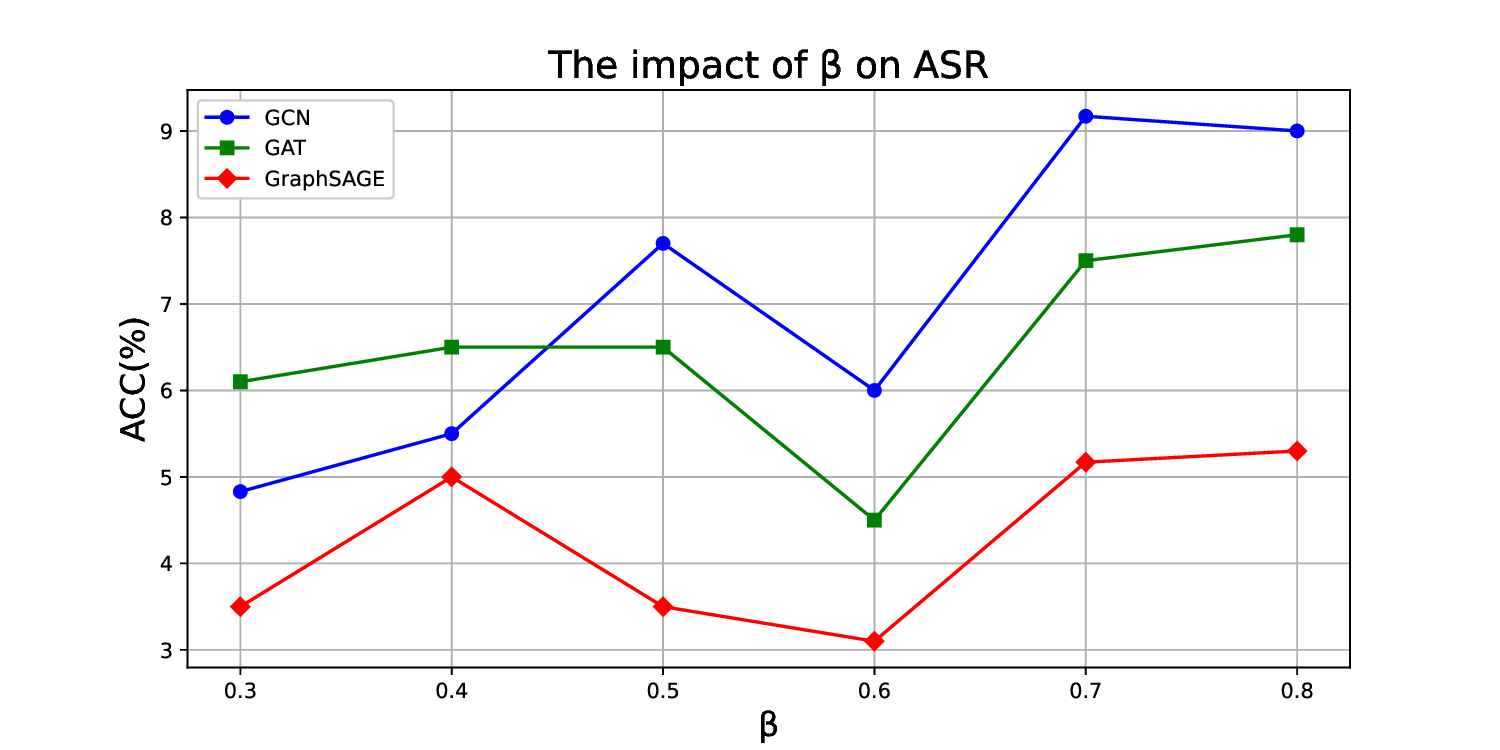}
\end{subfigure}
\hfill 
\begin{subfigure}[b]{0.47\textwidth}
  \centering
  \includegraphics[width=\linewidth]{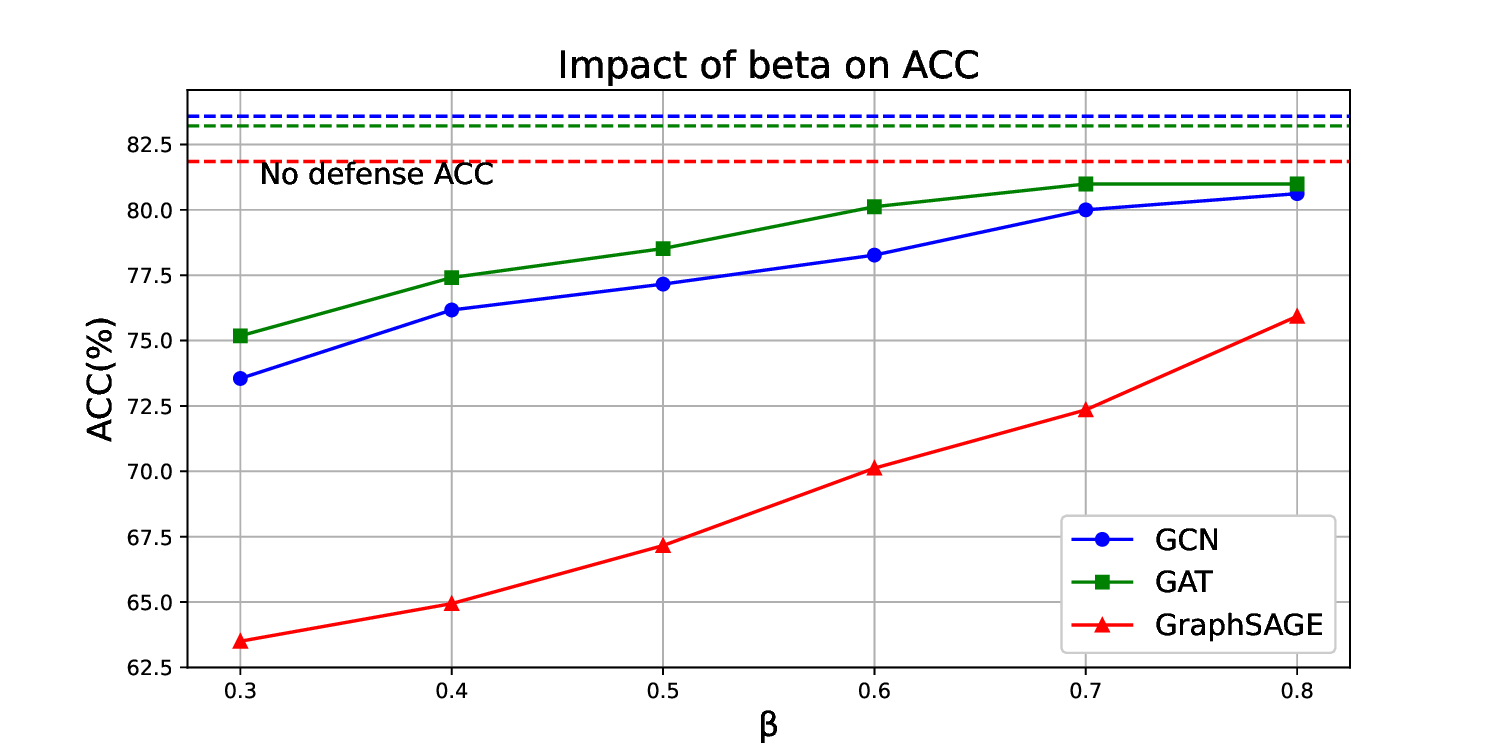}
\end{subfigure}
\hfill
\caption{The effect of \(\beta\) on ASR and ACC(Cora)}
\label{fig4}
\end{figure}

In the Cora dataset, both our method and the pruning method reduced the accuracy of the model. 
The decrease in accuracy is particularly pronounced in GraphSAGE, possibly because its aggregation strategy is more sensitive to pruning in low-degree nodes. Fig.~\ref{fig4} shows how the threshold \(\beta\) affects ACC and ASR (UGBA). This situation only occurs in small datasets. For medium and large datasets, the distribution of adversarial edges is nearly entirely distinct from that of the clean edges(see Fig.~\ref{fig2}), thus allowing for a wide range of threshold choices. Therefore, to ensure the accuracy of nodes with low degrees, we adopted a pruning threshold that adapts according to the node degree. Since the benefits of lowering the threshold are limited and the loss to accuracy is significant, we try to maintain a higher threshold when dealing with low-degree nodes. This strategy reflects a trade-off between defensive performance and accuracy.

\vspace{-0.4cm}

\subsection{Multiple Subgraph Insertion Experiments}
\vspace{-0.1cm}
To address more complex attack methods, we evaluate the performance of the defense algorithm in the context of multiple subgraph insertions in this section. 
The model used in this experiment is GCN. The experiment is divided into two parts: (i) attacking the model using TDGIA, and (ii) conducting multiple subgraph insertion attacks using the UGBA model. 
TDGIA is a type of adversarial attack based on subgraph insertion, aimed at reducing the model's ACC. Our defense method successfully maintained model accuracy against this attack, as shown on the left side of Fig.~\ref{fig5}.We have modified the UGBA algorithm to allow it to insert multiple triggers on the same node. On the right side, our defense effectively lowered the ASR for the UGBA attack, which involves inserting two or three subgraphs. This illustrates the efficacy of our defense in both preserving accuracy under direct adversarial challenges and reducing the success of complex backdoor attacks.

\begin{figure}[htp]
\centering
\begin{subfigure}[b]{0.47\textwidth}
  \centering
  \includegraphics[width=\linewidth]{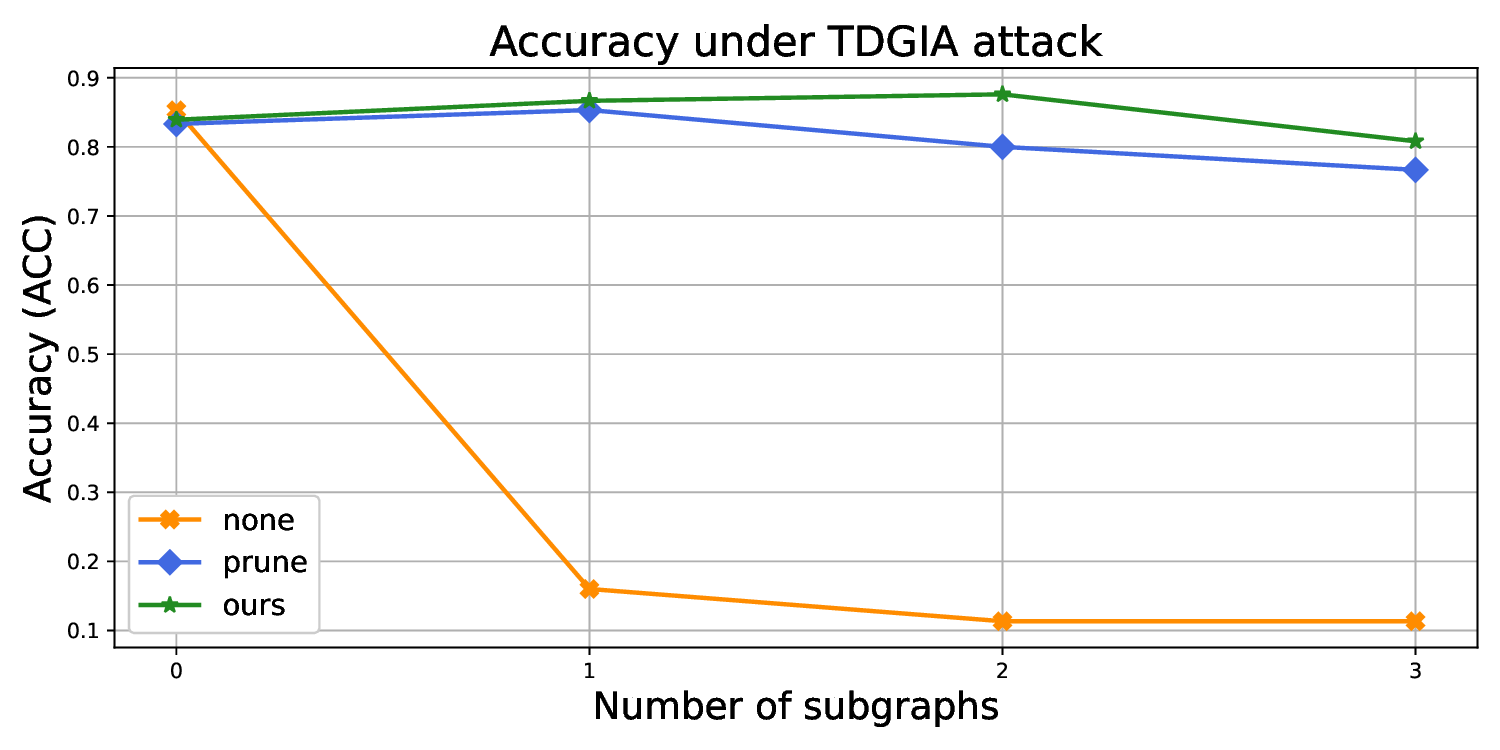}
\end{subfigure}
\hfill 
\begin{subfigure}[b]{0.47\textwidth}
  \centering
  \includegraphics[width=\linewidth]{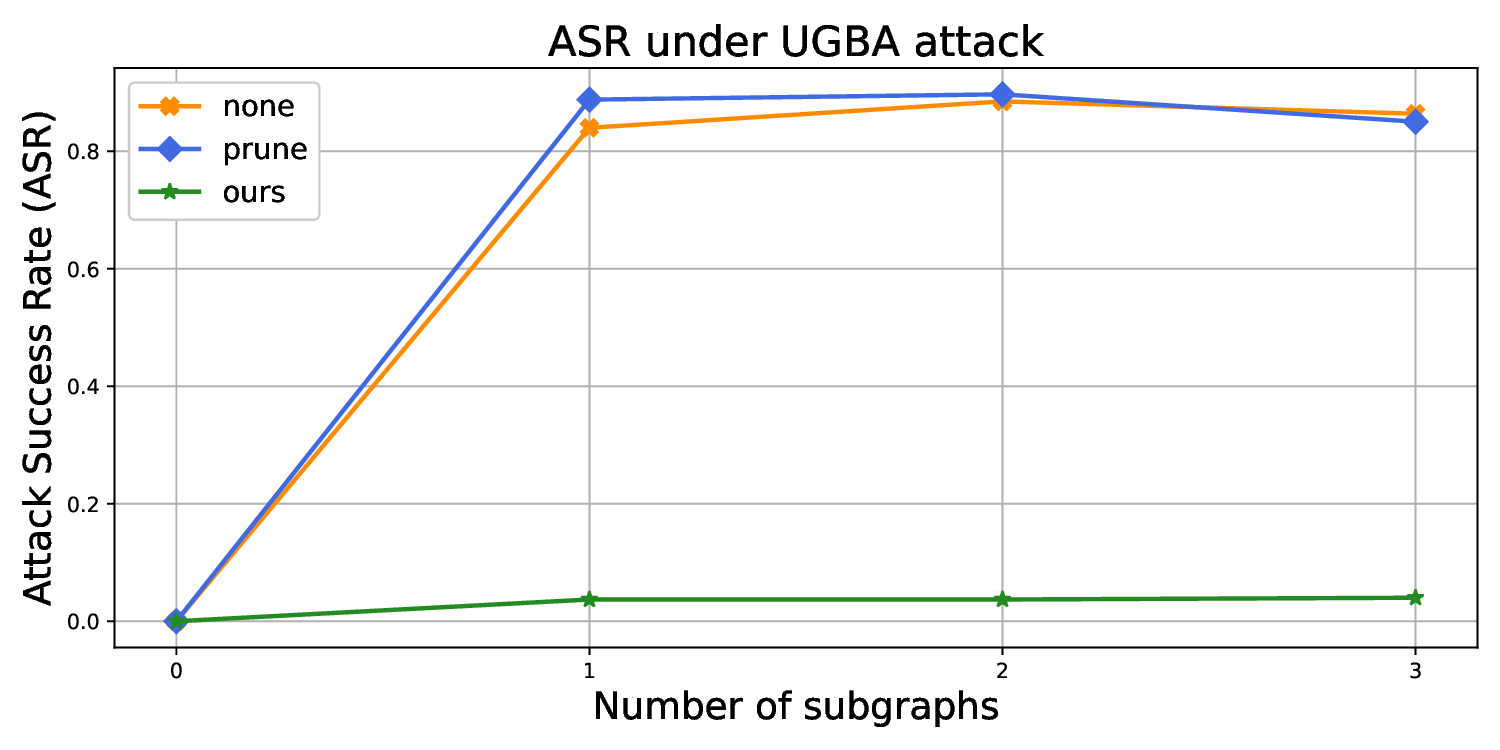}
\end{subfigure}
\hfill

\caption{Evaluation of multi-subgraph insertion attacks}
\label{fig5}
\end{figure}

\vspace{-0.4cm}

\subsection{Efficiency}
\vspace{-0.1cm}
By drawing on the ideas from GraphSAGE, we have managed to keep the time it takes to predict nodes in large datasets within a reasonable range through simplifying. This approach to simplifying subgraphs is based on the homophily of graphs, where adjacent nodes are often similar. In this section, we additionally introduce ogbn-arxiv, a large dataset with 160,000 nodes and 1,166,243 edges. Our experiments observed the performance of two datasets on the GCN model before and after optimization. After optimization, this method has reduced the runtime by 41.13$\%$ on Flickr(from 1.24s to 0.73s) and by 47.33$\%$ on ogbn-arxiv(From 4.62s to 2.43s).Although not comparable to method based on cosine similarity, this proves that our method is also feasible on large datasets.

\begin{figure}
\includegraphics[width=\textwidth]{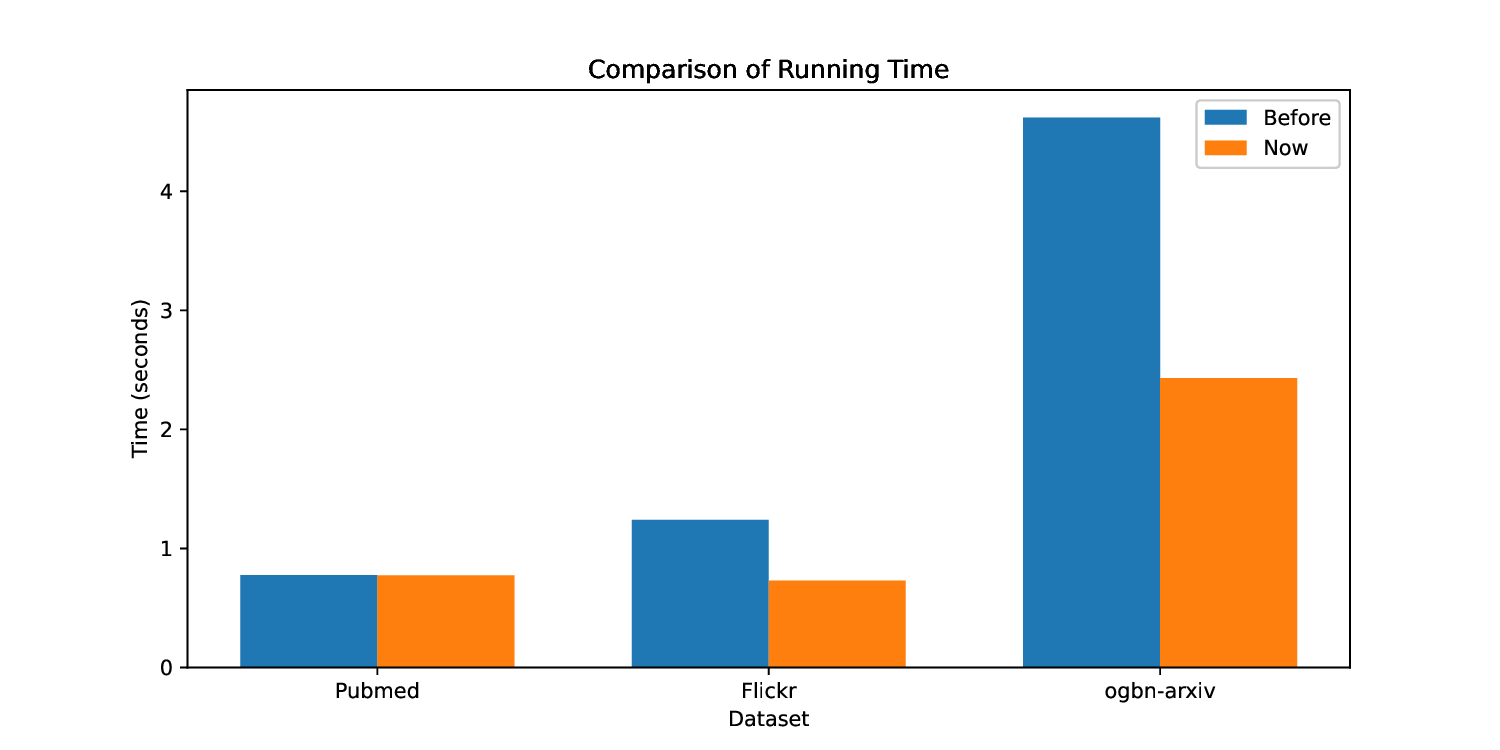}
\caption{Comparison of running time before and after optimization} \label{fig5}
\end{figure}

\vspace{-0.3cm}
\section{Conclusion and Future Work}
\vspace{-0.2cm}
In this paper, we propose a defense method based on the explainability of graph neural networks. This approach successfully defends against state-of-the-art backdoor attacks and is also applicable to other adversarial attacks based on subgraph insertion.
Furthermore, our defense has been optimized to address potential backdoor attacks involving the insertion of multiple triggers. In future work, we plan to study how to make triggers less obvious in explainability methods, which will effectively improve the stealth of attacks. In addition, it is also interesting to study the relationship between explainability tools and models.

\vspace{-0.2cm}

\bibliographystyle{splncs04}
\bibliography{esage}

\end{document}